\documentclass[10pt,conference]{IEEEtran}
\IEEEoverridecommandlockouts
\usepackage{cite}
\usepackage{amsmath,amssymb,amsfonts}
\usepackage{graphicx}
\usepackage{textcomp}
\usepackage{xcolor}
\usepackage{physics}
\usepackage{gensymb}
\usepackage{booktabs}
\usepackage{tabularx}
\usepackage{multirow}
\usepackage{optidef}
\usepackage[final]{microtype}
\usepackage{import}
\usepackage{xifthen}
\usepackage{pdfpages}
\usepackage{transparent}
\usepackage[linesnumbered,ruled,vlined]{algorithm2e}
\usepackage[flushleft]{threeparttable}
\usepackage[T1]{fontenc}
\usepackage[top=0.7in,bottom=1in,left=0.673in,right=0.64in]{geometry}
\usepackage{fancyhdr}
\newcommand{\ra}[1]{\renewcommand{\arraystretch}{#1}}
\def\BibTeX{{\rm B\kern-.05em{\sc i\kern-.025em b}\kern-.08em
    T\kern-.1667em\lower.7ex\hbox{E}\kern-.125emX}
}
\newcolumntype{Y}{>{\centering\arraybackslash}X}

\begin{document}

\title{Efficient Precoding for LEO Satellites: A Low-Complexity Matrix Inversion Method via Woodbury Matrix Identity and arSVD\\
\thanks{This research was funded by dtec.bw - the Digitalization and Technology Research Center of the Bundeswehr - which is financed by the European Union - NextGenerationEU.}
}
\author{\IEEEauthorblockN{Mohammad Momani, Thomas Delamotte and Andreas Knopp}
\IEEEauthorblockA{Institute of Information Technology, University of the Bundeswehr Munich, 85579 Neubiberg, Germany \\
Email: \{mohammad.momani, thomas.delamotte, andreas.knopp\}@unibw.de}
}

\maketitle
\thispagestyle{fancy}

\begin{abstract}
The increasing deployment of massive active antenna arrays in low Earth orbit (LEO) satellites necessitates computationally efficient and adaptive precoding techniques to mitigate dynamic channel variations and enhance spectral efficiency. Regularized zero-forcing (RZF) precoding is widely used in multi-user MIMO systems; however, its real-time implementation is limited by the computationally intensive inversion of the Gram matrix. In this work, we develop a low-complexity framework that integrates the Woodbury (WB) formula with adaptive randomized singular value decomposition (arSVD) to efficiently update the Gram matrix inverse as the satellite moves along its orbit. By leveraging low-rank perturbations, the WB formula reduces inversion complexity, while arSVD dynamically extracts dominant singular components, further enhancing computational efficiency. Monte Carlo simulations demonstrate that the proposed method achieves computational savings of up to 61\% compared to conventional RZF precoding with full matrix inversion, while incurring only a modest degradation in sum-rate performance. These results demonstrate that WB-arSVD offers a scalable and efficient solution for next-generation satellite communications, facilitating real-time deployment in power-constrained environments.
\end{abstract}

\section{Introduction}
The rapid deployment of low Earth orbit (LEO) satellite networks equipped with active antenna arrays is transforming global wireless communications, providing broadband access to remote areas, supporting internet of things (IoT) applications, and enabling real-time services in previously unreachable locations \cite{9502642}. However, ensuring high spectral efficiency in these networks is particularly challenging due to the rapid temporal variations in channel conditions and stringent computational constraints on satellite payloads \cite{6046158}.

Multi-user multiple-input multiple-output (MU-MIMO) precoding techniques, particularly regularized zero-forcing (RZF) precoding, are widely employed to optimize signal quality and mitigate inter-user interference in LEO-based communication systems. Although effective, these methods require computationally intensive matrix inversions, making them impractical for real-time adaptive processing on resource-constrained satellites. The primary bottleneck lies in the repeated inversion of the Gram matrix, which scales cubically with the number of users, significantly limiting scalability in large MIMO systems \cite{6554990, rusek2012scaling}. As LEO satellites traverse their orbits, continuously updating the matrix inverse becomes computationally prohibitive, necessitating more computationally efficient solutions.

Several approaches have been proposed to address the complexity of matrix inversion in large-scale MIMO precoding. The conjugate gradient (CG) method has been explored as an iterative solver to avoid direct inversion, reducing complexity by an order of magnitude compared to full RZF inversion. However, CG suffers from slow convergence in ill-conditioned channel matrices \cite{8780818}. Another method is the Neumann series expansion (NSE), which approximates the inverse using a truncated power series expansion. While this reduces complexity, the accuracy depends heavily on the number of expansion terms, and large iterations may approach the complexity of full inversion \cite{6777306}. Additionally, QR decomposition-based and other iterative methods such as Gauss-Seidel, successive over-relaxation (SOR), and Jacobi methods have been proposed for efficient inversion, with some approaches leveraging hardware acceleration. While QR decomposition provides numerical stability, it remains computationally expensive in power-limited satellite systems. Additionally, iterative methods often suffer from slow convergence and poor robustness to channel variations \cite{6554990}. More recently, efforts to simplify zero-forcing in extra-large MIMO (XL-MIMO) systems have leveraged plane-wave approximations and sub-array user grouping \cite{ribeiro2021low}. Although effective in spatially non-stationary, near-field terrestrial scenarios, these techniques are not designed for non-terrestrial networks (NTNs). These limitations highlight the need for efficient and adaptive solutions tailored to real-time, resource-constrained NTNs.

To address these challenges, we propose a novel computationally efficient framework that integrates the Woodbury (WB) formula with adaptive randomized singular value decomposition (arSVD) to accelerate the inversion of the Gram matrix. The WB formula enables incremental updates to the matrix inverse, reducing computational complexity by leveraging low-rank channel perturbations, while arSVD dynamically identifies dominant singular components, further optimizing efficiency. Our approach achieves an optimal balance between computational efficiency and system performance.

This paper makes three key contributions. First, it introduces a WB-based inverse update mechanism that significantly reduces the computational complexity of RZF precoding. Second, it incorporates arSVD to efficiently approximate low-rank channel perturbations. Finally, Monte Carlo simulations validate the proposed approach, demonstrating computational savings of up to 61\%, with a modest sum-rate degradation of up to 10\%.

The remainder of this paper is structured as follows: Section~\ref{sys_model} presents the system model. Section~\ref{WBIU} details the WB-based inverse update mechanism and its computational advantages. Section~\ref{arsvd} describes the implementation of arSVD for low-rank approximation. Section~\ref{prop_method} explains the proposed method. Section~\ref{sim_results} presents simulation results and performance analysis, and Section~\ref{conc} concludes the paper with key insights and future research directions.

\section{System Model}\label{sys_model}
We consider a MU-MIMO downlink (DL) communication system, where a satellite equipped with $N_t$ transmit antennas serves $K$ ground user terminals (UTs), each equipped with a single receive antenna. The satellite employs a fully connected hybrid beamforming structure with $N_{RF}$ radio frequency (RF) chains, satisfying $K \le N_{RF} \ll N_t$. Additionally, it is assumed that the UTs are capable of estimating and compensating for Doppler shifts.

The downlink channel matrix between the satellite's radiating elements and the UTs at time instant $t$ is denoted as
\begin{equation}
    \mathbf{H}(t) \in \mathbb{C}^{K \times N_t} = \left[\mathbf{h}_1(t), \mathbf{h}_2(t), \dots, \mathbf{h}_{K}(t)\right]^T,
\end{equation}
where $\mathbf{h}_n(t) \in \mathbb{C}^{N_t \times 1}$ represents the $n$-th UT channel vector. The channel follows a Rician fading model, given by
\begin{equation}
    \mathbf{h}_n(t) = \sqrt{\frac{K_R}{K_R+1}} \, \mathbf{h}_n^{\text{LOS}}(t) + \sqrt{\frac{1}{K_R+1}} \, \mathbf{h}_n^{\text{NLOS}}(t),
\end{equation}
where $K_R$ is the Rician factor, determining the relative power between the line-of-sight (LOS) component $\mathbf{h}_n^{\text{LOS}}(t)$ and the random normally distributed non-line-of-sight (NLOS) components $\mathbf{h}_n^{\text{NLOS}}(t)$. The deterministic LOS path $\mathbf{h}_n^{\text{LOS}}(t)$ is modeled as
\begin{equation}
    \mathbf{h}_n^{\text{LOS}}(t) = \gamma_n \, \widetilde{\mathbf{h}}_n(t) = \gamma_n \, \mathbf{a}\left(\theta_n(t), \phi_n(t)\right),
\end{equation}
where $\gamma_n$ is the $n$-th UT LOS gain including the distance-dependent free-space path loss and atmospheric losses and $\mathbf{a}\left(\theta_n(t), \phi_n(t)\right) = \frac{1}{\sqrt{N_t}} \left[ e^{j\boldsymbol{\kappa}\mathbf{r}_1}, e^{j\boldsymbol{\kappa}\boldsymbol{r}_2}, \dots , e^{j\boldsymbol{\kappa}\boldsymbol{r}_{N_t}} \right]^T$ is the array manifold vector of the $n$-th UT with the downlink azimuth and elevation angles of departure $(\theta_n(t), \phi_n(t))$. $\boldsymbol{\kappa} = \frac{2 \pi}{\lambda} \left[ \sin\theta_n(t) \cos\phi_n(t), \sin\theta_n(t) \sin\phi_n(t), \cos\theta_n(t) \right]$ is the wavevector, $\lambda$ is the wavelength and $\mathbf{r}_m = \left[x_m, y_m, z_m\right]^T$ is the position column vector of the $m$-th satellite radiating element. For brevity and clarity of notation, we omit explicit time dependence in subsequent expressions, noting that these variables remain functions of time.

The multi-user effective channel matrix can be written by incorporating analog beam steering at the satellite as
\begin{equation}
    \mathbf{H}_{\text{eff}} = \widetilde{\mathbf{H}} \, \mathbf{F}_{RF} \in \mathbb{C}^{K \times N_{RF}},
\end{equation}
where $\mathbf{\widetilde{H}} = \left[\mathbf{\widetilde{h}}_1, \mathbf{\widetilde{h}}_2, \dots, \mathbf{\widetilde{h}}_K \right]^T$ is the multi-user effective channel and $\mathbf{F}_{RF} \in \mathbb{C}^{N_t \times N_{RF}}$ is the analog beam steering matrix. Each UT’s DL analog beam steering vector is selected from a discrete Fourier transform (DFT) codebook containing a set of predefined angles, based on a best-power criterion \cite{10946029}. The effective channel matrix depends on the positions of the satellite and UTs, which we assume is known at the satellite. Under this formulation, the digital precoder can be computed without requiring channel estimation feedback from the UTs.

To mitigate multi-user interference, we employ RZF precoding and the digital precoding matrix $\mathbf{F}_{BB} \in \mathbb{C}^{N_{RF} \times K}$ can be written in terms of the effective channel as
\begin{equation} \label{eq:RZF}
    \mathbf{F}_{BB} = \mathbf{H}_{\text{eff}}^H\left(\mathbf{H}_{\text{eff}} \, \mathbf{H}^H_{\text{eff}} + \alpha \mathbf{I}_K\right)^{-1} = \mathbf{H}_{\text{eff}}^H \, \mathbf{A}^{-1},
\end{equation}
where $\alpha \ge 0$ is the regularization parameter, $\mathbf{I}_K$ is the identity matrix of size $K \times K$ and $\mathbf{A}$ is the Gram matrix. To enforce the total transmit power constraint, $\mathbf{F}_{BB}$ is scaled such that $\|\mathbf{F}_{RF} \mathbf{F}_{BB} \|_F^2 = P_t$, where $P_t$ denotes the satellite's total transmit power. The primary computational challenge in \eqref{eq:RZF} lies in the inversion of the Gram matrix.

As the satellite moves in orbit, the channel matrix $\mathbf{H}_{\text{eff}}$ undergoes continuous perturbations, rendering frequent recomputation of $\mathbf{A}^{-1}$ computationally prohibitive.
To address this challenge, our proposed method avoids full inversion by utilizing
\begin{itemize}
    \item arSVD to approximate the perturbations efficiently.
    \item WB identity to update $\mathbf{A}^{-1}$ using low-rank modifications.
\end{itemize}

\section{WB-Based Inverse Update}\label{WBIU}
If $\mathbf{H}_{\text{eff}}$ changes due to small perturbations as the satellite moves in orbit, the updated effective channel matrix can be written as
\begin{equation}
    \mathbf{H}_{\text{eff}} ^\prime = \mathbf{H}_{\text{eff}} + \Delta \mathbf{H}_{\text{eff}},
\end{equation}
\noindent where $\Delta \mathbf{H}_{\text{eff}}$ represents the perturbations in the effective channel caused by satellite movement. The updated Gram matrix then becomes

\begin{equation} \label{A_prime}
    \mathbf{A}^\prime = \left(\mathbf{H}_{\text{eff}} + \Delta \mathbf{H}_{\text{eff}}\right) \left(\mathbf{H}_{\text{eff}} + \Delta \mathbf{H}_{\text{eff}}\right)^H + \alpha \mathbf{I}_K.
\end{equation}
Expanding \eqref{A_prime}, we obtain

\begin{equation} \label{A_prime2}
    \mathbf{A}^\prime = \mathbf{H}_{\text{eff}} \mathbf{H}_{\text{eff}}^H + \mathbf{H}_{\text{eff}} \Delta \mathbf{H}_{\text{eff}}^H + \Delta \mathbf{H}_{\text{eff}} \mathbf{H}_{\text{eff}}^H + \Delta \mathbf{H}_{\text{eff}} \Delta \mathbf{H}_{\text{eff}} ^H + \alpha \mathbf{I}_K.
\end{equation}
Simplifying further
\begin{equation}\label{A_prime2_2}
    \mathbf{A}^\prime = \mathbf{A}  + \underbrace{\mathbf{H}_{\text{eff}} \Delta \mathbf{H}_{\text{eff}}^H + \Delta \mathbf{H}_{\text{eff}} \mathbf{H}_{\text{eff}}^H + \Delta \mathbf{H}_{\text{eff}} \Delta \mathbf{H}_{\text{eff}} ^H}_\text{$\Delta \mathbf{A}$}.
\end{equation}
From \eqref{A_prime2_2}, the updated Gram matrix $\mathbf{A}^\prime$, reflecting the satellite's new orbital position, can be represented using the current Gram matrix $\mathbf{A}$ and additional correction terms. These extra terms capture how the current effective channel correlates with any new perturbations introduced by the satellite's movement, forming what we refer to as the Gram matrix correction $\Delta \mathbf{A}$.

If $\Delta \mathbf{A}$ serves as a low-rank correction to $\mathbf{A}$, and if it can be represented in the form of $\mathbf{U}_r\mathbf{\Sigma}_r\mathbf{V}_r^H$, then we can rewrite the updated Gram matrix in \eqref{A_prime2_2} as
\begin{equation} \label{A_prime3}
    {\mathbf{A}^\prime} = \mathbf{A} + \mathbf{U}_r\mathbf{\Sigma}_r\mathbf{V}_r^H,
\end{equation}
and its inverse as
\begin{equation} \label{A_prime3_3}
    {\mathbf{A}^\prime}^{-1} = \left(\mathbf{A} + \mathbf{U}_r\mathbf{\Sigma}_r\mathbf{V}_r^H\right)^{-1},
\end{equation}
where $\mathbf{A}$ is rank-$K$ and $\mathbf{U}_r$ and $\mathbf{V}^H_r$ are $K\times r$ orthogonal matrices, and $\mathbf{\Sigma}_r$ is an $r \times r$ diagonal matrix of non-zero singular values $\sigma_j$, with $r$ denoting the rank of the correction matrix. This is known as the WB matrix identity and provides an efficient method for updating a matrix inverse following a low-rank modification, without recomputing the full inverse, provided that $r < K$. Instead, it involves inverting an auxiliary, smaller $r \times r$ matrix \cite{doi:10.1137/1.9781421407944}. In other words, it provides a direct expression for the inverse of the updated Gram matrix ${\mathbf{A}^\prime}^{-1}$ in terms of an already-computed inverse ${\mathbf{A}}^{-1}$ and $\Delta \mathbf{A}$ as
\begin{equation}\label{eq_WB}
    \begin{aligned}
        &\left(\mathbf{A} + \mathbf{U}_r\mathbf{\Sigma}_r\mathbf{V}_r^H\right)^{-1}\\
        & = \mathbf{A}^{-1} - \mathbf{A}^{-1}\mathbf{U}_r(\mathbf{\Sigma}_r^{-1} + \mathbf{V}_r^H\mathbf{A}^{-1}\mathbf{U}_r)^{-1}\mathbf{V}_r^H \mathbf{A}^{-1}.
    \end{aligned}
\end{equation}
Equation \eqref{eq_WB} holds under the assumption that both $\mathbf{A}$ and $(\mathbf{\Sigma}_r^{-1} + \mathbf{V}_r^H\mathbf{A}^{-1}\mathbf{U}_r)$ are non-singular.

When $\Delta \mathbf{A}$ is a low-rank correction ($r \ll K$), the WB identity can significantly reduce the computational complexity of computing the next Gram matrix inverse compared to carrying out a full matrix inversion from scratch. By starting from a previously computed inverse $\mathbf{A}^{-1}$, we avoid recalculating the new full inverse and instead focus on inverting the much smaller matrix $\left(\mathbf{\Sigma}_r^{-1} + \mathbf{V}_r^H\mathbf{A}^{-1}\mathbf{U}_r\right)^{-1}$, whose size depends on the rank of the perturbation. This low-rank assumption holds because changes to $\Delta \mathbf{H}_{\text{eff}}$ typically involve small updates to the effective channel, affecting only a limited subset of its dimensions as shown in \cite{10946029}. Consequently, the WB identity emerges as a particularly efficient approach for handling matrix inversions in these dynamic scenarios aboard power-limited satellite platforms.

\subsection{WB Inverse Update Computational Complexity}\label{sec:WB_comp}
The standard complexity for computing a full matrix inversion of a $K\times K$ matrix is $\mathcal{O}\left(K^3\right)$.

Now, for updating the inverse of a matrix using the WB formula in \eqref{eq_WB}, we
\begin{itemize}
    \item[1.] Compute $\mathbf{A}^{-1}\mathbf{U}_r$: $\mathcal{O}\left(K^2 r\right)$.
    \item[2.] Compute $\mathbf{V}_r^H\mathbf{A}^{-1}\mathbf{U}_r$: $\mathcal{O}\left(K r^2\right)$.
    \item[3.] Invert the matrix $\left(\mathbf{\Sigma}_r^{-1} + \mathbf{V}_r^H\mathbf{A}^{-1}\mathbf{U}_r\right)^{-1}$: $\mathcal{O}\left(r^3\right)$.
    \item[4.] Compute the product of $\mathbf{A}^{-1}\mathbf{U}_r$ with the inverse from the previous step. Then multiply that with $\mathbf{V}_r^H\mathbf{A}^{-1}$: $\mathcal{O}\left(K r^2 + K^2 r\right)$.
    \item[5.] Finally, subtract this result from $\mathbf{A}^{-1}$: $\mathcal{O}\left(K^2\right)$.
\end{itemize}
Thus, the total computational complexity of the WB formula is approximated as $\mathcal{O}\left(r^3 + K r^2 + K^2 r + K^2\right)$. These results are also summarized in Table~\ref{tab_complex}.
\begin{table}[t]
    \centering
    \caption{Computational complexity of full matrix inversion and WB inverse update.}
    \label{tab_complex}
    \ra{1.1}
    \begin{tabularx}{\columnwidth}{XX} \toprule
    \textbf{Method} & \textbf{Complexity} \\ \midrule
    Full matrix inversion & $\mathcal{O}\left(K^3\right)$ \vspace{0.2cm} \\
    Inverse update using WB identity  & $\mathcal{O}\left(K^2 + K^2 r + r^3 + r^2 K\right)$ \\ \bottomrule
    \end{tabularx}
    \vspace{-0.3cm}
\end{table}

\section{Finding a Low-Rank Approximation of $\mathbf{\Delta A}$}\label{arsvd}
To employ the WB formula for updating a previously computed inverse $\mathbf{A}^{-1}$ after a low-rank modification by $\Delta\mathbf{A}$, we need to approximate $\Delta\mathbf{A}$ as a low-rank matrix. A common method is truncated singular value decomposition (SVD). The full SVD of $\Delta\mathbf{A}$ is given as
\begin{equation}
    \Delta\mathbf{A} = \mathbf{U}\mathbf{\Sigma}\mathbf{V}^H.
\end{equation}
where $\mathbf{U}$, $\mathbf{V}$ and $\mathbf{\Sigma}$ are $K \times K$ matrices.

To approximate $\Delta\mathbf{A}$ as a low-rank matrix, we retain only the largest $r$ singular values (and their corresponding singular vectors) such that the retained singular values capture at least $\eta$ of the total energy (squared Frobenius norm) of $\Delta\mathbf{A}$. Mathematically

\begin{equation}
    \frac{\sum_{j=1}^{r} \sigma_j^2}{\sum_{j=1}^{\text{rank}(\Delta\mathbf{A})} \sigma_j^2} \ge \eta \;\;\;\;\;\; (0 < \eta < 1).
\end{equation}

A full SVD of a $K\times K$ matrix typically requires $\mathcal{O}(K^3)$ operations, which is on the same order as directly inverting $\mathbf{A}^\prime$. Consequently, relying on frequent full SVD computations to obtain a lower-rank approximation negates the benefit of maintaining and updating the inverse using the WB formula. The added complexity of performing the full SVD negates the computational savings intended. As a result, we explore a more efficient iterative approach from the family of randomized SVD algorithms \cite{doi:10.1137/090771806}.

\subsection{Adaptive Randomized SVD (arSVD)}
The goal is to approximate the matrix $\mathbf{\Delta A}$ by finding a low-rank subspace that preserves a fraction $\eta$ of its Frobenius norm energy. Let $E_{\text{target}} = \eta  \| \mathbf{\Delta A} \|_F^2$ denote the target energy to be retained. The algorithm is initialized with a small rank estimate $k_\text{init}$ and an oversampling parameter $p$, and the total subspace dimension is defined as $d = \min(k_0 + p, K)$, where $k_0$ is the working rank estimate, initially set equal to $k_\text{init}$ for the first iteration. At each iteration, a random Gaussian matrix $\mathbf{\Omega} \in \mathbb{C}^{K \times d}$ is generated, and the sketch $\mathbf{Y} = \mathbf{\Delta A} \mathbf{\Omega}$ is formed. An orthonormal basis $\mathbf{Q} \in \mathbb{C}^{K \times d}$ for the column space of $\mathbf{Y}$ is then obtained via thin QR decomposition. The matrix $\mathbf{\Delta A}$ is projected onto this subspace by computing $\mathbf{B} = \mathbf{Q}^H \mathbf{\Delta A}$. The SVD $\mathbf{B} = \hat{\mathbf{U}} \hat{\mathbf{\Sigma}} \hat{\mathbf{V}}^H$ is then computed, and the cumulative energy $\sum_{j=1}^r \sigma_j^2$ is evaluated for $r \leq d$. If $\sum_{j=1}^r \sigma_j^2 \geq E_{\text{target}}$ for some $r$, the approximation $\mathbf{\Delta A} \approx \mathbf{U}_r \mathbf{\Sigma}_r \mathbf{V}_r^H$ is returned, where $\mathbf{U}_r$, $\mathbf{\Sigma}_r$, and $\mathbf{V}_r$ are the truncated SVD components. Otherwise, the working rank estimate $k_0$ is doubled, and the process repeats with a newly generated subspace of higher dimension (with oversampling applied through $d$). The iteration continues until either the energy condition is satisfied or a predefined maximum number of iterations is reached. This adaptive expansion balances computational effort with approximation accuracy, ensuring that effort is focused on capturing the dominant modes of $\mathbf{\Delta A}$. The complete procedure is summarized in Algorithm~\ref{alg:adaptive-randsvd}.

\begin{algorithm}[!bp]
    \caption{Adaptive Randomized SVD}
    \label{alg:adaptive-randsvd}
    \KwIn{Matrix $\mathbf{\Delta A}\in\mathbb{C}^{K\times K}$, energy threshold $\eta\in(0,1]$, initial rank $k_{\text{init}}$, oversampling parameter $p$, max iterations $I_{\max}$.}
    \KwOut{Approx.\ SVD factors $\mathbf{U}_r,\mathbf{\Sigma}_r,\mathbf{V}_r$ and $k_{\mathrm{est}}$ s.t.\ $\|\mathbf{\Delta A}-\mathbf{U}_r\mathbf{\Sigma}_r\mathbf{V}_r^H\|_F^2 \approx (1-\eta)\|\mathbf{\Delta A}\|_F^2$.}
    \BlankLine
    $E_{\text{target}}\gets \eta\cdot\|\mathbf{\Delta A}\|_F^2$\;
    $k_0\gets k_{\text{init}}$\;
    \For{$i=1$ \KwTo $I_{\max}$}{
      $d\gets \min(k_0 + p, K)$\;
      Draw $\mathbf{\Omega}\in\mathbb{C}^{K\times d}$ with i.i.d.\ $\mathcal{C}\mathcal{N}(0,1)$ entries\;
      $\mathbf{Y}\gets \mathbf{\Delta A}\,\mathbf{\Omega}$\;
      $\mathbf{Q}\gets \mathrm{orth}(\mathbf{Y})$\;
      $\mathbf{B}\gets \mathbf{Q}^H\,\mathbf{\Delta A}$\;
      $[\hat{\mathbf{U}},\hat{\mathbf{\Sigma}},\hat{\mathbf{V}}]\gets \mathrm{svd}(\mathbf{B})$\;
      $\mathbf{U}_{\mathrm{approx}}\gets \mathbf{Q}\,\hat{\mathbf{U}}$\;
      Let $\sigma_j$ be the $j$-th singular value of $\hat{\mathbf{\Sigma}}$\;
      Find smallest $r$ s.t.\ $\sum_{j=1}^r\sigma_j^2 \ge E_{\text{target}}$\;
      \eIf{such $r$ exists}{
        $k_{\mathrm{est}}\gets r$\;
        $\mathbf{U}_r\gets \mathbf{U}_{\mathrm{approx}}(:,1\!:\!r)$\;
        $\mathbf{\Sigma}_r\gets \hat{\mathbf{\Sigma}}(1\!:\!r,1\!:\!r)$\;
        $\mathbf{V}_r\gets \hat{\mathbf{V}}(:,1\!:\!r)$\;
        \Return{$(\mathbf{U}_r,\mathbf{\Sigma}_r,\mathbf{V}_r, k_{\mathrm{est}})$}\;
      }{
        $k_0\gets 2\,k_0$\;
      }
    }
    \BlankLine
    $k_{\mathrm{est}}\gets d$\;
    $\mathbf{U}_r\gets \mathbf{U}_{\mathrm{approx}}(:,1\!:\!d)$\;
    $\mathbf{\Sigma}_r\gets \hat{\mathbf{\Sigma}}(1\!:\!d,1\!:\!d)$\;
    $\mathbf{V}_r\gets \hat{\mathbf{V}}(:,1\!:\!d)$\;
    \Return{$(\mathbf{U}_r,\mathbf{\Sigma}_r,\mathbf{V}_r, k_{\mathrm{est}})$}\;
\end{algorithm}
This iterative strategy avoids the need for a full SVD by limiting each iteration to multiplying $\mathbf{\Delta A}$ by a random matrix of the current subspace dimension, followed by an SVD on the resulting smaller matrix. When comparing the computational cost of approximating $\mathbf{\Delta A}$ via a full SVD versus arSVD, a substantial reduction is observed in favor of the latter. A detailed complexity comparison is provided in Table~\ref{tab_complex_svd}. It is worth noting that our complexity estimate does not explicitly account for the iterative nature of the arSVD algorithm or the oversampling parameter $p$. This simplification is justified by the fact that arSVD is inherently parallelizable—each random projection step is independent—and our Monte Carlo simulations show that convergence is typically achieved within two to three iterations when the working rank $k$ is doubled at each step. Furthermore, the typical choice of $p$ is small enough that its impact on overall complexity remains negligible.
 
\begin{table}[t]
    \centering
    \caption{Computational complexity of finding a low-rank approximation of $\mathbf{\Delta A}$ using full SVD and ar-SVD.}
    \label{tab_complex_svd}
    \ra{1.1}
    \begin{tabularx}{\columnwidth}{XX} \toprule
    \textbf{Method} & \textbf{Complexity} \\ \midrule
    Full SVD & $\mathcal{O}\left(K^3\right)$ \vspace{0.2cm} \\
    arSVD  & $\mathcal{O}\left(K^2 r + r^2 K\right)$ \\ \bottomrule
    \end{tabularx}
\end{table}
\section{Proposed Method}\label{prop_method}
Our proposed method combines the WB formula for updating the inverse of the Gram matrix as the satellite moves along its orbit with the arSVD algorithm for approximating the perturbation matrix $\mathbf{\Delta A}$ as a low-rank factorization of the form $\mathbf{\Delta A} \approx \mathbf{U}_r \, \mathbf{\Sigma}_r \, \mathbf{V}^H_r$.

\begin{figure}[t]
    \includegraphics[width=\columnwidth]{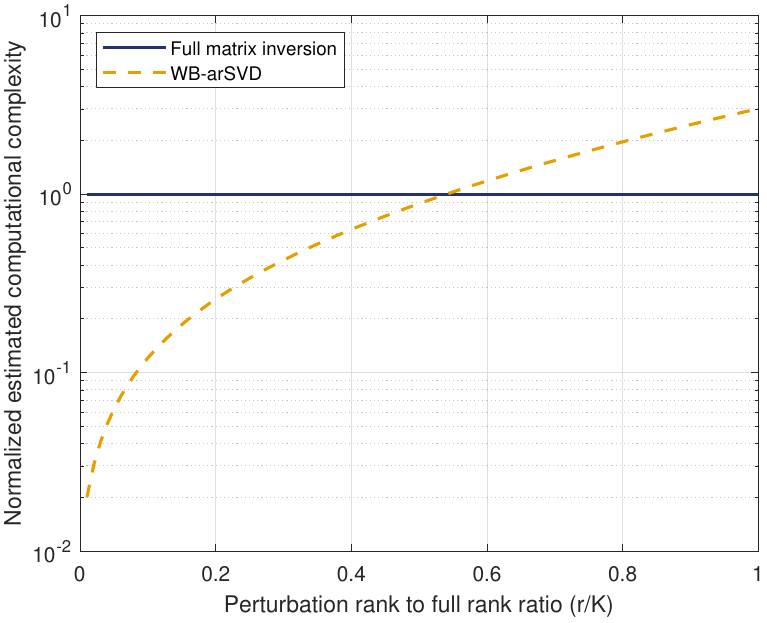}
    \caption{Normalized estimated computational complexity from Table~\ref{tab_complex_total} for fixed $K$ and variable $r \to K$. $K = 100$. Y-axis is in log scale.}
    \label{fig_compl}
\end{figure}

By combining the complexity results from Table~\ref{tab_complex} and Table~\ref{tab_complex_svd}, we compare the overall computational complexity of the proposed method against that of directly inverting the Gram matrix, as summarized in Table~\ref{tab_complex_total}. To provide further insight, we contrast the complexity of full matrix inversion with that of updating a precomputed inverse using the WB-arSVD method, assuming a fixed matrix dimension $K$ and varying perturbation rank $r$. The corresponding results are illustrated in Fig.~\ref{fig_compl}. From this figure, three distinct cases can be identified:
\begin{itemize}
    \item When ($r \ll K$), the WB-arSVD approach is highly efficient compared to full matrix inversion and significantly accelerates the computation of the updated matrix inverse. 
    \item When $r/K \approx 0.5$, the WB-arSVD approach remains slightly more efficient than full matrix inversion.
    \item As $r$ increases (particularly past about $r = 0.5 K$), the $r^3$ and $r^2 K$ terms become comparable to the $K^3$ complexity of direct inversion. Eventually, as $r \to K$, the WB-arSVD method becomes several times more expensive than direct full matrix inversion.
\end{itemize}
In conclusion, the WB-arSVD approach proves particularly effective when the perturbation to the Gram matrix ($\mathbf{\Delta A}$) is of low rank ($r \ll K$). However, when the perturbation does not exhibit a significantly reduced rank relative to the original matrix, direct full matrix inversion becomes the more practical option to avoid unnecessary computational overhead. Based on these observations, we propose the method outlined in Algorithm~\ref{alg:proposedMethod}. In the proposed procedure, the arSVD is first computed on $\mathbf{\Delta A}$, and the ratio $r/K$ is evaluated. If $r/K \leq 0.5$, the inverse of the updated Gram matrix is computed efficiently using the WB formula. Otherwise, a full matrix inversion of $(\mathbf{A}^\prime)^{-1}$ is performed. 

It is important to note that the empirical breakpoint at $r/K \approx 0.5$ may depend on the specific scenario and system characteristics, including hardware capabilities and implementation details. Future work could explore the development of an adaptive threshold that dynamically balances computational efficiency and approximation accuracy.

\begin{table}[t]
    \centering
    \caption{Computational complexity of full matrix inversion and the combination of WB-arSVD.}
    \label{tab_complex_total}
    \ra{1.1}
    \begin{tabularx}{\columnwidth}{XX} \toprule
    \textbf{Method} & \textbf{Complexity} \\ \midrule
    Full matrix inversion & $\mathcal{O}\left(K^3\right)$ \vspace{0.2cm} \\
    WB-arSVD  & $\mathcal{O}\left(K^2 + K^2 r + r^3 + r^2 K\right)$ \\ \bottomrule
    \end{tabularx}
\end{table}

\begin{algorithm}[bh!]
    \caption{Proposed Method}
    \label{alg:proposedMethod}
    \eIf{$\mathbf{A}^{-1}$ exists}{%
      Compute $\mathbf{A}^\prime$ and $\mathbf{\Delta A}$ from \eqref{A_prime2_2}\;
      Perform arSVD on $\mathbf{\Delta A}$ using Algorithm~\ref{alg:adaptive-randsvd}\;
      \eIf{$k_{est}/K \le 0.5$}{%
        Apply WB formula in \eqref{eq_WB} to compute $(\mathbf{A}^\prime)^{-1}$\;
      }{%
        Compute $(\mathbf{A}^\prime)^{-1}$ by direct matrix inversion\;
      }
    }{%
      Compute $\mathbf{A}^{-1}$ by direct matrix inversion\;
    }
  \end{algorithm}
\section{Simulation Results \& Discussion}\label{sim_results}
To evaluate system performance, we use the Shannon-capacity–based system sum-rate as our key performance indicator (KPI), defined as
\begin{equation}\label{opt_prob}
    R_{sum} = \sum_{n = 1}^K  \operatorname*{log_2} \left( 1 + \frac{\left| \mathbf{h}^H_n \, \mathbf{F}_{RF} \, \mathbf{f}_{BB,n}\right| ^2}{\sum_{i \neq n} \left| \mathbf{h}^H_n \, \mathbf{F}_{RF} \, \mathbf{f}_{BB,i}\right| ^2 \, + \, \sigma_n^2} \right),
\end{equation}
where $\sigma_n^2$ is the additive white Gaussian noise (AWGN) variance. The simulation parameters are summarized in Table~\ref{tab_sim_param}.
\begin{figure}[t!]
    \includegraphics[width=\columnwidth]{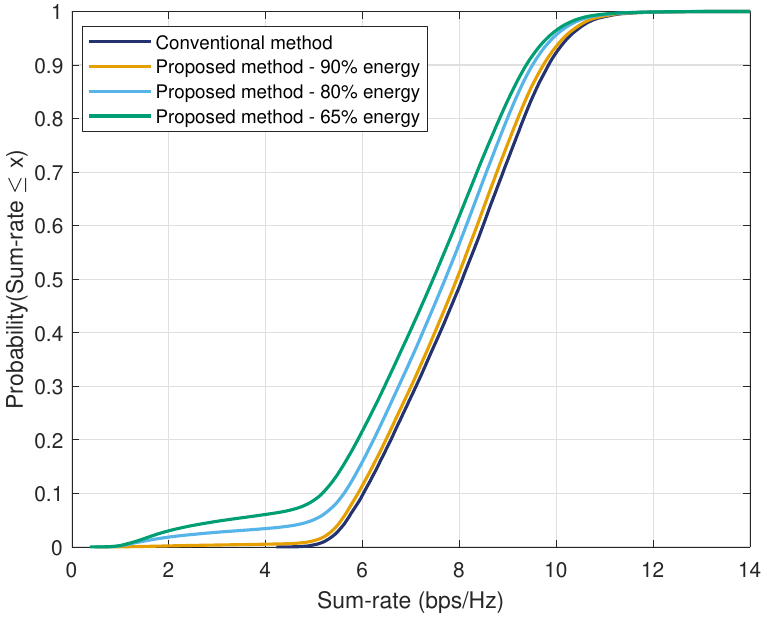}
    \caption{Empirical cumulative distribution function (ECDF) of the Shannon system sum-rate for RZF digital precoding with different energy thresholds $\eta$. The digital precoding matrix is updated every 50 ms.}
    \label{fig_RZF_WB}
\end{figure}
\begin{table}[t]
    \centering
    \caption{Simulation parameters.}
    \label{tab_sim_param}
    \ra{1.15}
\begin{threeparttable}
    \begin{tabularx}{\columnwidth}{XX}\toprule
    \textbf{Parameter}  & \textbf{Value}  \\ \midrule
    Carrier frequency  & 18 GHz \\
    Orbital height & 600 km \\
    Antenna array & $16 \times 16$\\
    Antenna element spacing  & $\lambda/2$ \\
    $N_{RF}$      & 16 \\
    $P_t$ & 20 dBW \\ 
    UT antenna gain  & 39.7 dBi \tnote{1}\\ 
    UT distribution  & Uniform \\
    $K$ & up to 16 \\ 
    Number of active UTs per time/frequency frame per beam & \multirow{2}{*}{1} \\ 
    Simulated satellite pass duration & 120 s\\  
    Number of Monte Carlo runs & 500 \\
    Digital precoding update rate & 20 times per second \\
    $p$ & 1 \\
    $k_{init}$ & 2\\
    $K_R$ & 10 dB\\ \bottomrule
    \end{tabularx}%
    \begin{tablenotes}
        \item[1] Value is adopted from \cite{3gpp3rdGenerationPartnership2023}.
      \end{tablenotes}
\end{threeparttable}
\vspace{-0.6cm}
\end{table}

Fig.~\ref{fig_RZF_WB} presents the system sum-rate obtained from Monte Carlo simulations, comparing the conventional RZF precoding approach, which computes a full matrix inversion at each simulation snapshot, with the proposed method outlined in Algorithm~\ref{alg:proposedMethod}. The proposed method is evaluated at various target energy retention levels for the arSVD computation (Algorithm~\ref{alg:adaptive-randsvd}).

The results demonstrate that the proposed method incurs moderate performance degradation compared to the conventional approach, with the decrease in sum-rate proportional to the retained energy level. This reduction is primarily attributed to the iterative nature of the arSVD process, which, while efficiently updating the low-rank representation of $\mathbf{\Delta A}$, may introduce small approximation errors.

Notably, even at lower energy retention levels (down to 65\% of the original matrix energy), the dominant features of the effective channel perturbations are well preserved across successive snapshots. This is reflected by the moderate sum-rate degradation observed in our simulations. One possible explanation is that the considered scenario involves a relatively small number of active simultaneous UTs, with $K \leq N_{\text{RF}} = 16$. In such cases, a 65\% energy threshold is sufficient to retain the majority of the matrix structure, as the dominant singular values capture most of the energy. This observation is consistent with the principle that when the energy is highly concentrated in a few singular values, low-rank approximations yield high accuracy.

Furthermore, Table~\ref{tab_saving} highlights the significant computational savings achieved by the proposed method. Compared to the conventional full-matrix inversion approach, the proposed method reduces computational complexity by approximately 30\% at 90\% energy retention, 47\% at 80\% energy retention, and 61\% at 65\% energy retention. These reductions not only lower computational costs but also substantially decrease processing time, making the method well-suited for real-time implementation in power-constrained satellite systems. Moreover, these efficiency gains support the hypothesis that Gram matrix perturbations induced by satellite movement are inherently low-rank, with only a small number of singular values significantly impacting system performance.

The higher sum-rate degradation observed at 80\% and 65\% energy retention (5.8\% and 9.5\% loss, respectively) suggests that, while WB-arSVD effectively preserves the dominant singular components, the smaller singular values contribute to residual multi-user interference. Further investigation into the role of the regularization parameter $\alpha$ in mitigating these effects will be pursued in future work.

\begin{table}[t]
    \centering
    \caption{Average computational savings and sum-rate (SR) degradation of the proposed method compared to the conventional method, based on Monte Carlo simulations and Table~\ref{tab_complex_total}.}
    \label{tab_saving}
    \ra{1.1}
    \begin{tabularx}{\columnwidth}{XYY} \toprule
    \textbf{Method} & \textbf{Comp. savings (\%)} & \textbf{SR degradation (\%)} \\ \midrule
    Conventional method & \,\,\,$0.0\%$ & $0.0\%$ \vspace{0.1cm} \\
    Proposed method for various $\eta$ &  \\ 
    \quad \quad @ 90\% & $30.6\%$ & $1.6\%$ \\
    \quad \quad @ 80\% & $47.7\%$ & $5.8\%$ \\
    \quad \quad @ 65\% & $61.2\%$ & $9.5\%$ \\ \bottomrule
    \end{tabularx}
    \vspace{-0.5cm}
\end{table}

Real-time deployment critically depends on balancing the digital precoding update rate with computational efficiency. Frequent updates enhance sum-rate performance \cite{10946029, 10437228} and help preserve the low-rank structure of effective channel perturbations—an essential condition that enables the proposed algorithm to update the Gram matrix inverse with minimal overhead. However, high update rates may exceed the processing capabilities of current LEO satellite hardware. In contrast, less frequent updates reduce the computational burden but risk degrading sum-rate performance and increasing the perturbation rank, thereby diminishing the benefits of the proposed method. Consequently, selecting an appropriate update rate requires carefully balancing hardware constraints, performance targets, and the computational gains enabled by the proposed approach.
\vspace{-0.22cm}
\section{Conclusion}\label{conc}
In this paper, we developed a computationally efficient framework for accelerating RZF precoding in LEO satellite communications by integrating the WB formula with arSVD. The proposed method efficiently updates the Gram matrix inverse under low-rank channel perturbations, significantly reducing computational complexity. Monte Carlo simulations demonstrate that the approach achieves approximately 30\%–61\% computational savings compared to conventional full-matrix inversion, while incurring minimal to moderate degradation in system sum-rate performance. Nevertheless, the proposed method has limitations. When channel perturbations become high-rank, computational savings diminish, and full matrix inversion may become preferable. Moreover, the approximation errors introduced by arSVD lead to minor performance degradation, motivating further optimization. Future work will focus on hardware-optimized implementations of WB-arSVD on FPGA and GPU platforms to enable real-time deployment. These developments will help extend the applicability of the proposed approach to next-generation LEO satellite networks, 6G MIMO systems, and real-time beamforming applications.
\vspace{-0.05cm}
\bibliographystyle{ieeetr}
\bibliography{lib}
\end{document}